\documentclass[12pt,letterpaper]{article}
\usepackage[a4paper, total={7in, 10in}]{geometry}

\usepackage{graphicx}
\usepackage{helvet}
\usepackage{authblk}
\usepackage{hyperref}
\usepackage{amsmath} 
\usepackage{amssymb} 
\usepackage{orcidlink} 
\usepackage[super,comma,sort&compress]  
   {natbib}
\usepackage[right]{lineno} \linenumbers

\usepackage{booktabs}
 \usepackage{multirow}

\makeatletter
\renewcommand{\maketitle}{\bgroup\setlength{\parindent}{0pt}
\begin{flushleft}
  \textbf{\@title}
  
  \@author
\end{flushleft}\egroup}
\makeatother

\title{EFGPP: Exploratory framework for genotype-phenotype prediction}
\date{}

\author[1,2]{Muhammad Muneeb}
\author[1,2,\orcidlink{0000-0003-2948-2413},*]{David B. Ascher}

\affil[1]{School of Chemistry and Molecular Biology, The University of Queensland, Brisbane, 4067, Australia}
\affil[2]{Computational Biology and Clinical Informatics, Baker Heart and Diabetes Institute, Melbourne, 3004, Australia}

\affil[*]{Correspondence: d.ascher@uq.edu.au}

\begin{document}

\maketitle

\section*{SUMMARY}
Predicting complex phenotypes from human genetic data is increasingly constrained not by data availability, but by how best to prioritize and combine heterogeneous sources of signal. Here, we present \textsc{EFGPP}, a systematic framework for generating, ranking, and integrating heterogeneous genetic data representations for genotype--phenotype prediction. Using UK Biobank data from 733 individuals, we evaluated migraine prediction by combining genotype-derived features, principal components, clinical and metabolomic covariates, and PRS generated from two migraine GWAS and three depression GWAS using \textsc{PLINK}, \textsc{PRSice-2}, \textsc{AnnoPred}, and \textsc{LDAK-GWAS}. The best individual representation achieved a test AUC of 0.644, while structured multimodal integration improved performance to 0.688 in a migraine-focused configuration and 0.663 in a cross-trait configuration incorporating depression-derived inputs. No single genetic modality matched the covariates-only baseline (AUC 0.639), although genotype features outperformed PRS in isolation and depression-derived PRS remained competitive. These results position \textsc{EFGPP} as a reproducible proof-of-concept framework for systematic data prioritization and multimodal integration in genotype--phenotype prediction.

\section*{KEYWORDS}


genotype-phenotype prediction, genetics, gwas, machine learning, polygenic risk scores 

\section*{INTRODUCTION}
Genotype-phenotype prediction involves predicting traits and diseases based on genetic datasets \cite{Visscher2017,Medvedev2022}. It has diverse applications, including understanding disease mechanisms \cite{RN6695}, personalized medicine \cite{LISTGARTEN2013,Dong2022}, case-control classification, disease treatment responses \cite{RN5710}, and exploring the interplay between genetic diseases \cite{RN8133,Muneeb2022}.

\begin{figure}[!ht]
    \centering
    \includegraphics[width=1\textwidth]{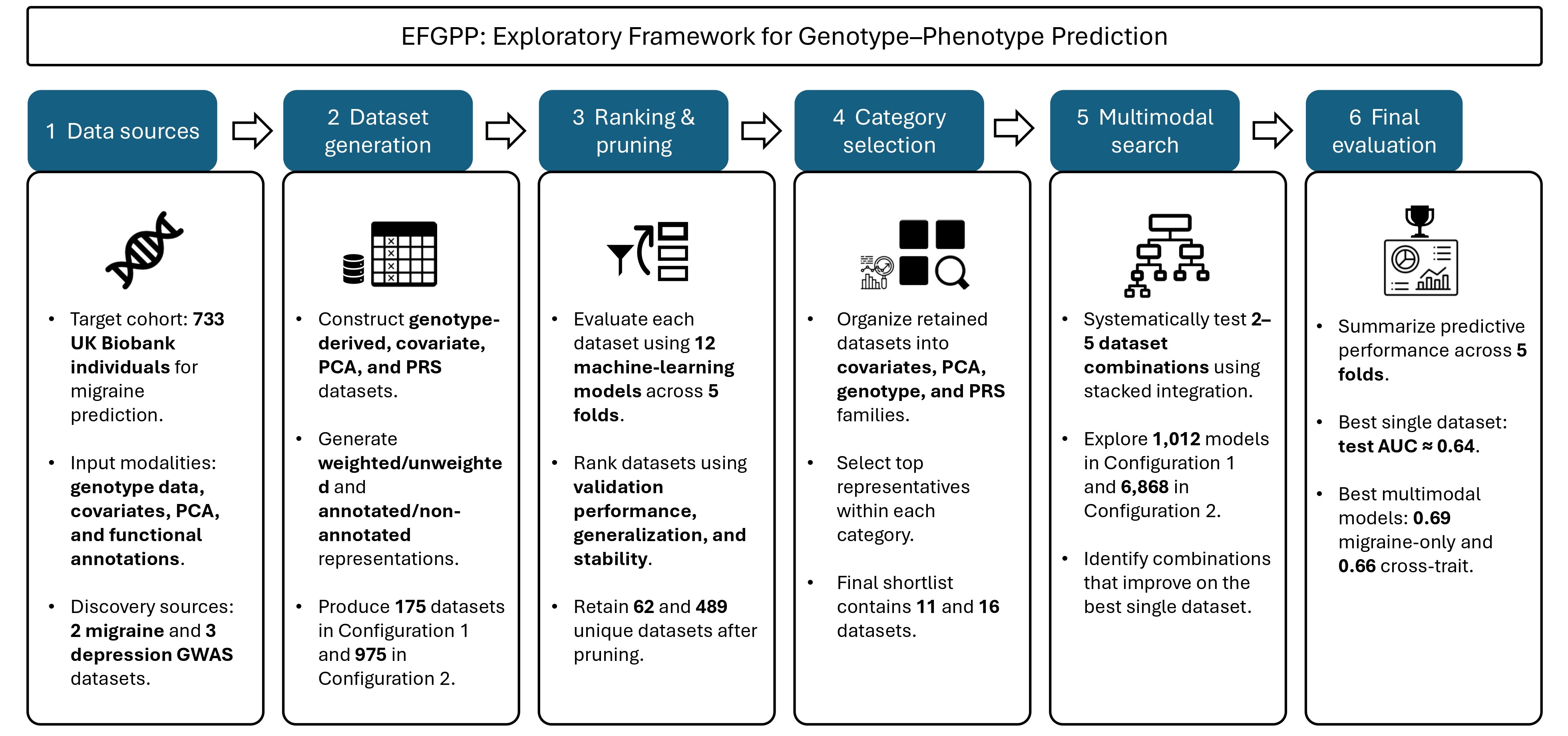}
    \caption{\textbf{Overview of the \textsc{EFGPP} workflow.} The framework proceeds through six main stages. \textbf{(1) Data sources:} heterogeneous inputs are assembled from the target cohort and external discovery resources, including genotype data, covariates, principal components, functional annotations, and GWAS summary statistics. \textbf{(2) Dataset generation:} multiple individual datasets are constructed, including genotype-derived, covariate, PCA, and PRS-based representations. \textbf{(3) Ranking and pruning:} each dataset is benchmarked across machine-learning models and ranked according to predictive performance, generalization, and stability, while redundant representations are removed. \textbf{(4) Category selection:} the strongest retained datasets are grouped into representative families such as covariates, PCA, genotype, and PRS. \textbf{(5) Multimodal search:} shortlisted datasets are systematically combined to evaluate whether integrated representations improve predictive performance beyond any single modality. \textbf{(6) Final evaluation:} the best individual and multimodal models are compared using cross-validated performance metrics to quantify the gain from structured data integration.}
    \label{workflow}
\end{figure}

Data from various biological, environmental \cite{Hunter2005}, and phenotypic dimensions are gathered and incorporated into a coherent framework for robust predictive modeling \cite{Guo2023}. These data sources include, but are not limited to, genotype data obtained by selecting top single nucleotide polymorphisms (SNPs) using p-value thresholding \cite{Fadista2016} from genome-wide association studies (GWAS). Covariates encompass medical conditions, metabolite levels, sex, and other demographic or clinical variables that may influence phenotypic expression \cite{McCaw2022}. Principal component analysis (PCA) is applied to the genotype data to account for population stratification and reduce dimensionality \cite{Price2006}. Functional annotations (FA) provide insights into the biological relevance of genetic variants, assisting in prioritizing SNPs for prediction models \cite{Torkamani2011}. GWAS offers summary statistics highlighting associations between genetic variants and traits, serving as a foundation for identifying significant SNPs \cite{Tam2019}. Polygenic risk scores (PRS) \cite{Torkamani2018,Khera2018}, derived from GWAS data, aggregate the effects of multiple genetic variants to estimate an individual's genetic predisposition to specific traits or diseases. These data sources are combined to create a feature set, and machine learning (ML) or deep learning (DL) algorithms are employed for phenotype prediction \cite{Guo2023,Sehrawat2023}. Researchers have further enhanced prediction performance by integrating multiple GWAS \cite{Ritchie2015,Garreta2021} for one phenotype, leveraging GWAS from related phenotypes \cite{2018mtag} and populations \cite{Ishigaki2022}, and combining multiple PRS \cite{Truong2024}. For instance, combining GWAS data from migraine and depression enables the development of multi-trait PRS, capturing shared genetic architectures between related conditions.
To address the wide variety of data representations and integration strategies available for genotype--phenotype prediction, we developed \textsc{EFGPP}, an exploratory framework that systematically generates, prioritizes, and integrates heterogeneous datasets. The framework first constructs individual datasets from covariates, genotype-derived matrices (weighted or unweighted, annotated or non-annotated), principal components, and polygenic risk scores (PRS) generated using \textsc{PLINK} \cite{Purcell_2007}, \textsc{PRSice-2} \cite{Choi_2019}, \textsc{AnnoPred} \cite{Zheng_2024}, and \textsc{LDAK-GWAS} \cite{Zhang_2021}. These individual datasets are then benchmarked and ranked according to predictive performance and stability, after which the strongest representatives from each category are carried forward for multimodal integration. The framework supports multiple GWAS from the same or related phenotypes, incorporates functional annotation, and enables structured comparison of genetic and non-genetic inputs within a unified modelling workflow. In this way, \textsc{EFGPP} serves as a practical decision framework for identifying which data representations and combinations are most informative for case--control prediction under limited-sample conditions. Figure~\ref{workflow} summarizes the overall workflow of \textsc{EFGPP}.

Despite this progress, there remains limited practical guidance on how to systematically evaluate, rank, and combine these heterogeneous inputs in a reproducible way, particularly in modestly sized datasets where exhaustive exploration can quickly introduce redundancy and overfitting. The field now has many possible genetic representations, PRS tools, and source GWAS, but no simple framework for deciding which combinations are worth carrying forward for a given phenotype under realistic data limitations. This gap is especially consequential for complex traits with moderate polygenic signal, where the marginal contribution of each data modality is small and selection decisions have a disproportionate impact on out-of-sample performance. Migraine provides a useful test case in which to address this problem directly. It is a common and clinically heterogeneous neurological disorder with a substantial polygenic component, and it frequently co-occurs with depression. Beyond epidemiological comorbidity, genetic studies have identified overlapping polygenic architecture and shared loci between migraine and psychiatric traits including depression~\cite{Ligthart2014}, suggesting that cross-trait information may contain predictive signal not captured by migraine-only models. At the same time, related-trait integration is not guaranteed to improve performance, making migraine and depression a stringent setting in which to test a framework designed to prioritize informative inputs rather than simply accumulate features. Here, we present \textsc{EFGPP}, an exploratory framework for genotype--phenotype prediction that addresses this practical model-development problem: given multiple GWAS, PRS tools, genotype representations, functional annotations, and covariates, which combinations are most worth carrying forward for a given phenotype? Framed in this way, \textsc{EFGPP} is best viewed as a decision framework for systematic data prioritization in genotype--phenotype modelling and as a foundation for larger-scale benchmarking, external validation, and future biological refinement.

\section*{RESULTS}

\subsection*{EFGPP constructs and prioritizes heterogeneous genotype--phenotype representations}

We first used EFGPP to generate and evaluate heterogeneous representations for migraine prediction, including covariates, principal components, genotype-derived matrices, and PRS derived from migraine and depression GWAS. In the migraine-focused configuration, 174 candidate datasets were generated and reduced to 62 non-redundant representations after similarity-based pruning. In the expanded cross-trait configuration, 975 candidate datasets were reduced to 489 non-redundant representations. Following fold-wise model benchmarking, validation-based filtering, and composite-score ranking, 11 datasets from the migraine-focused configuration and 16 datasets from the cross-trait configuration were carried forward for multimodal integration. This staged reduction illustrates the main purpose of EFGPP: to make large representation spaces tractable while preserving modality diversity for downstream prediction.



Across individual datasets, the strongest single-representation performance was observed for a weighted genotype dataset derived from \texttt{migraine.gz} with 50 annotated SNPs (\texttt{snps\_annotated\_50}) evaluated using logistic regression, achieving a mean training AUC of 0.655758 ($\pm$ 0.02147), validation AUC of 0.637858 ($\pm$ 0.086888), and test AUC of 0.643899 ($\pm$ 0.143235). The most stable individual representation in Configuration~1 was a PRS dataset generated from \texttt{migraine.gz} using LDAK-GWAS and evaluated with Naive Bayes, yielding a training AUC of 0.670031 ($\pm$ 0.027755), validation AUC of 0.638314 ($\pm$ 0.161106), and test AUC of 0.575402 ($\pm$ 0.028343). In Configuration~2, several datasets showed stable behaviour, including the corresponding LDAK-GWAS PRS representation and a genotype representation based on \texttt{depression\_3.gz} with 5000 unweighted SNPs, but none exceeded the best individual performance observed for the compact weighted migraine-derived genotype representation.

Datasets with mean validation AUC exceeding 0.6 were retained for downstream prioritization and combination testing; using the machine-learning benchmark, 49 and 355 datasets passed this threshold in Configurations~1 and~2, respectively. Following composite-score ranking within each category, 11 datasets were selected for Configuration~1 and 16 for Configuration~2, as shown in Table~\ref{tab:datasets}. The remaining datasets that passed the validation threshold are available on GitHub (\texttt{Configuration1/ResultsTop10.csv} and \texttt{Configuration2/ResultsTop10.csv}), and the final selected datasets for both configurations are provided on GitHub (\texttt{Configuration1/best\_datasets.csv} and \texttt{Configuration2/best\_datasets.csv}).

\begin{table}[!ht]
\centering
\caption{\textbf{Final selected datasets for both configurations.} Covariates and PCA appeared in both configurations. UW indicates genotype data not weighted by the GWAS file. W indicates genotype data weighted by the GWAS file. For genotype data, categories follow the format \texttt{Genotype\_\{Weighted/Unweighted\}\_\{Annotated/NotAnnotated\}\_\{GWASFile\}}. PRS datasets follow the format \texttt{PRS\_\{GWASFile\}\_\{PRSTool\}}. Configuration~2 included more GWAS files, resulting in five additional PRS datasets.}
\label{tab:datasets}
\small
\renewcommand{\arraystretch}{1.05}
\begin{tabular}{p{0.15\linewidth}p{0.38\linewidth}p{0.38\linewidth}}
\toprule
\textbf{Dataset type} & \textbf{Configuration~1 categories} & \textbf{Configuration~2 categories} \\
\midrule
Covariates & Covariates & Covariates \\
Genotype & Genotype\_UW\_annotated\_migraine & Genotype\_UW\_annotated \\
Genotype & Genotype\_UW\_not\_annotated\_migraine & Genotype\_UW\_not\_annotated \\
Genotype & Genotype\_W\_annotated\_migraine & Genotype\_W\_annotated \\
Genotype & Genotype\_W\_not\_annotated\_migraine & Genotype\_W\_not\_annotated \\
PCA & PCA & PCA \\
PRS & PRS\_migraine\_2\_AnnoPred & PRS\_depression\_1\_AnnoPred \\
PRS & PRS\_migraine\_2\_LDAK-GWAS & PRS\_migraine\_AnnoPred \\
PRS & PRS\_migraine\_AnnoPred & PRS\_migraine\_2\_AnnoPred \\
PRS & PRS\_migraine\_LDAK-GWAS & PRS\_depression\_2\_LDAK-GWAS \\
PRS & PRS\_migraine\_PLINK & PRS\_migraine\_LDAK-GWAS \\
PRS & -- & PRS\_migraine\_2\_LDAK-GWAS \\
PRS & -- & PRS\_depression\_2\_PRSice-2 \\
PRS & -- & PRS\_depression\_2\_PLINK \\
PRS & -- & PRS\_depression\_1\_PLINK \\
PRS & -- & PRS\_migraine\_PLINK \\
\bottomrule
\end{tabular}
\end{table}

\subsection*{Single-modality analyses show that no genetic representation is sufficient alone}

\begin{figure}[!ht]
    \centering
    \includegraphics[width=0.95\textwidth]{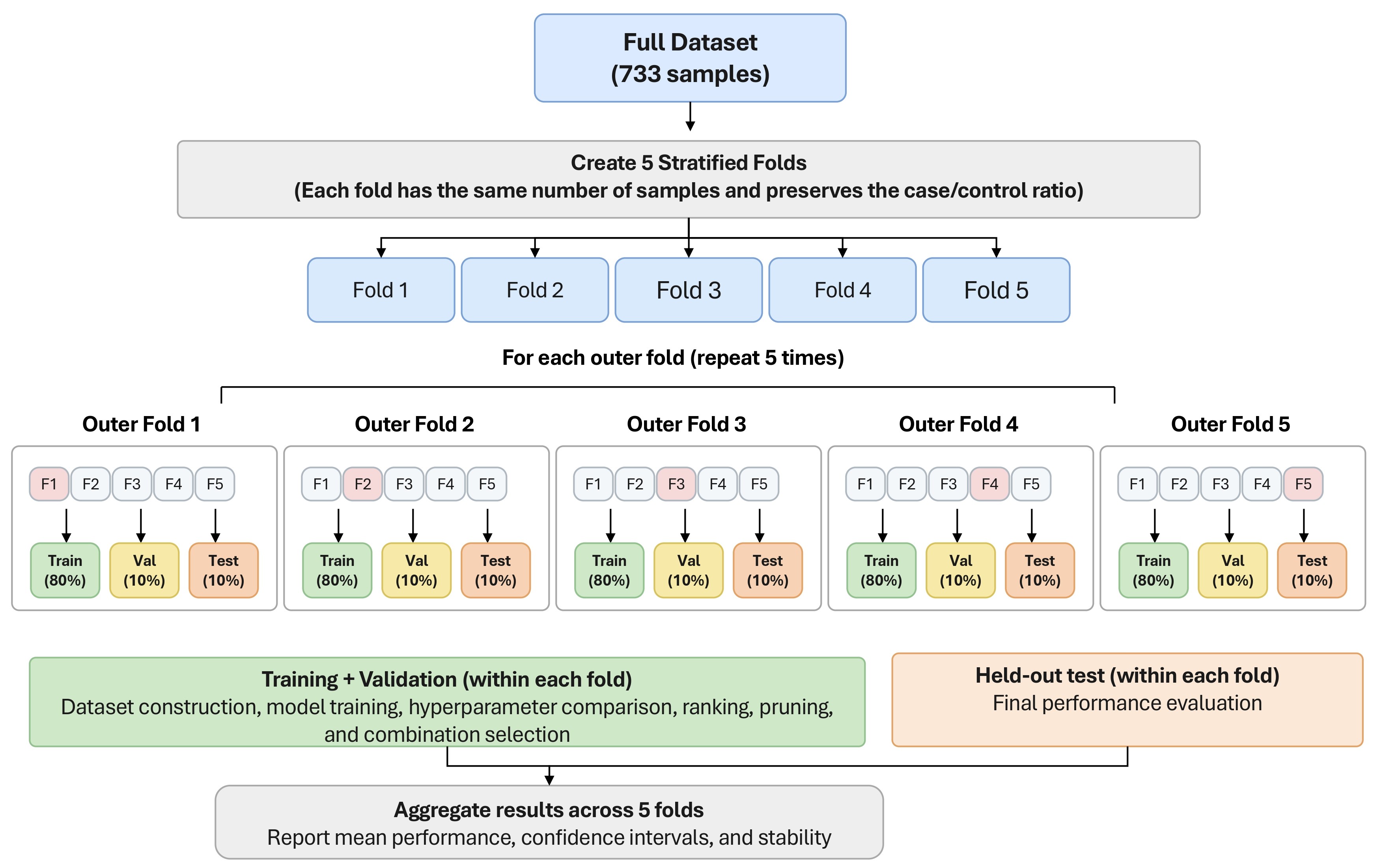}
    \caption{\textbf{Fold-wise evaluation design used in EFGPP.} The full dataset was partitioned into five stratified folds while preserving the migraine case--control ratio. For each fold, training, validation, and held-out test subsets were defined. Training and validation subsets were used for dataset construction, model fitting, hyperparameter comparison, ranking, pruning, and multimodal combination selection. The held-out test subset was reserved exclusively for final fold-level performance evaluation and was not used to guide dataset selection, model choice, or combination search. Final results were aggregated across all five folds to summarize predictive performance, confidence intervals, generalizability, and stability.}
    \label{fig:evaluation_design}
\end{figure}

To quantify the independent predictive contribution of each data modality, we conducted an ablation analysis across all individual datasets generated by EFGPP, averaging performance across five cross-validation folds and reporting 95\% confidence intervals (Table~\ref{tab:ablation}). The covariates-only model, comprising 135 NMR metabolomic biomarkers and comorbid condition indicators, achieved the highest test AUC among all single-modality configurations (AUC\,=\,0.639, 95\% CI [0.494, 0.784]), establishing a strong non-genetic baseline. No genetic modality alone matched this baseline.
To reduce information leakage during this large representation search, all data-dependent processing and model-selection steps were performed within each fold before held-out test evaluation (Figure~\ref{fig:evaluation_design}). Training and validation subsets were used for dataset construction, model fitting, hyperparameter comparison, ranking, pruning, and multimodal combination selection, whereas held-out test subsets were reserved exclusively for final fold-level performance evaluation.

Among PRS models derived from migraine GWAS, PLINK performed best (AUC\,=\,0.521, 95\% CI [0.443, 0.600], $\Delta=-0.118$), followed by LDAK-GWAS (AUC\,=\,0.497, $\Delta=-0.142$), PRSice-2 (AUC\,=\,0.476, $\Delta=-0.163$), and AnnoPred (AUC\,=\,0.471, $\Delta=-0.168$). Notably, PRS derived from depression GWAS performed comparably or better than migraine-derived PRS across all four tools, with LDAK-GWAS-depression achieving the strongest result among all PRS configurations (AUC\,=\,0.564, 95\% CI [0.520, 0.608], $\Delta=-0.075$), suggesting that depression-derived polygenic signal captures partially overlapping genetic architecture relevant to migraine risk, consistent with documented comorbidity and shared genetic loci between the two traits~\cite{Ligthart2014}.

PCA alone was a weak predictor (AUC\,=\,0.543, $\Delta=-0.096$), consistent with population-structure components not directly encoding trait-specific risk. Genotype-based features were more competitive than PRS in isolation. Unweighted annotated and weighted annotated genotype matrices derived from migraine GWAS both achieved a mean test AUC\,=\,0.605 (95\% CI [0.397, 0.812], $\Delta=-0.034$), and depression-GWAS-derived unweighted non-annotated genotype matrices reached AUC\,=\,0.620 (95\% CI [0.535, 0.705], $\Delta=-0.019$), the closest any single genetic modality came to matching the covariate baseline. Across all categories, the gap relative to covariates ranged from $\Delta=-0.019$ for the best genotype configuration to $\Delta=-0.168$ for the weakest PRS configuration, confirming that no single genetic representation is sufficient and motivating the multimodal integration strategy central to EFGPP.

\begin{table}[!ht]
\centering
\caption{\textbf{Single-modality ablation analysis for migraine prediction.} For each data category, the best-performing individual dataset was identified by mean test AUC across five cross-validation folds. Performance is reported as mean~$\pm$~SD with 95\% confidence interval (CI) computed using the $t$-distribution. $\Delta$ denotes the difference in mean test AUC relative to the covariates-only baseline. GWAS source indicates whether migraine or depression summary statistics were used to derive the representation.}
\label{tab:ablation}
\small
\resizebox{\textwidth}{!}{%
\begin{tabular}{llcccccc}
\toprule
\textbf{Category} & \textbf{GWAS source} & \textbf{Train AUC} & \textbf{Val AUC} & \textbf{Test AUC} & \textbf{SD} & \textbf{95\% CI} & \textbf{$\Delta$ vs cov.} \\
\midrule
\multicolumn{8}{l}{\textit{Non-genetic baselines}} \\
Covariates only      & --        & 0.9995 & 0.7438 & 0.6390 & 0.1169 & [0.494, 0.784] & 0.000 \\
PCA only             & --        & 0.8655 & 0.7803 & 0.5428 & 0.1461 & [0.361, 0.724] & $-$0.096 \\
\midrule
\multicolumn{8}{l}{\textit{PRS derived from migraine GWAS}} \\
PRS PLINK            & Migraine  & 0.9990 & 0.7849 & 0.5213 & 0.0634 & [0.443, 0.600] & $-$0.118 \\
PRS LDAK-GWAS        & Migraine  & 0.9673 & 0.7918 & 0.4967 & 0.1198 & [0.348, 0.645] & $-$0.142 \\
PRS PRSice-2         & Migraine  & 0.5256 & 0.7520 & 0.4762 & 0.0677 & [0.392, 0.560] & $-$0.163 \\
PRS AnnoPred         & Migraine  & 0.9354 & 0.7484 & 0.4711 & 0.0896 & [0.360, 0.582] & $-$0.168 \\
\midrule
\multicolumn{8}{l}{\textit{PRS derived from depression GWAS}} \\
PRS LDAK-GWAS        & Depression & 0.9720 & 0.7292 & 0.5641 & 0.0353 & [0.520, 0.608] & $-$0.075 \\
PRS PRSice-2         & Depression & 0.9733 & 0.7278 & 0.5546 & 0.1292 & [0.394, 0.715] & $-$0.084 \\
PRS AnnoPred         & Depression & 0.8079 & 0.7071 & 0.5286 & 0.1514 & [0.341, 0.717] & $-$0.110 \\
PRS PLINK            & Depression & 0.8592 & 0.7214 & 0.5188 & 0.1336 & [0.353, 0.685] & $-$0.120 \\
\midrule
\multicolumn{8}{l}{\textit{Genotype matrices derived from migraine GWAS}} \\
Unweighted annotated & Migraine  & 0.9798 & 0.8141 & 0.6045 & 0.1668 & [0.397, 0.812] & $-$0.034 \\
Weighted annotated   & Migraine  & 0.9798 & 0.8141 & 0.6045 & 0.1668 & [0.397, 0.812] & $-$0.034 \\
Unweighted not annotated & Migraine & 0.9671 & 0.6980 & 0.5701 & 0.1094 & [0.434, 0.706] & $-$0.069 \\
Weighted not annotated   & Migraine & 0.9671 & 0.6980 & 0.5701 & 0.1094 & [0.434, 0.706] & $-$0.069 \\
\midrule
\multicolumn{8}{l}{\textit{Genotype matrices derived from depression GWAS}} \\
Unweighted not annotated & Depression & 0.9309 & 0.7474 & 0.6197 & 0.0684 & [0.535, 0.705] & $-$0.019 \\
Unweighted annotated     & Depression & 0.9149 & 0.7566 & 0.6095 & 0.1078 & [0.476, 0.743] & $-$0.030 \\
Weighted not annotated   & Depression & 0.9309 & 0.7474 & 0.6197 & 0.0684 & [0.535, 0.705] & $-$0.019 \\
Weighted annotated       & Depression & 0.9149 & 0.7566 & 0.6095 & 0.1078 & [0.476, 0.743] & $-$0.030 \\
\bottomrule
\end{tabular}%
}
\end{table}

\subsection*{Data-generation choices materially affect performance and stability}

For each configuration, exploratory data analysis was performed to assess how data-generation parameters influenced individual-dataset performance. These analyses informed dataset prioritization; the primary claims of the study are based on the controlled single-modality and multimodal analyses reported above and below.

\begin{figure}[!ht]
\centering
\includegraphics[width=\textwidth]{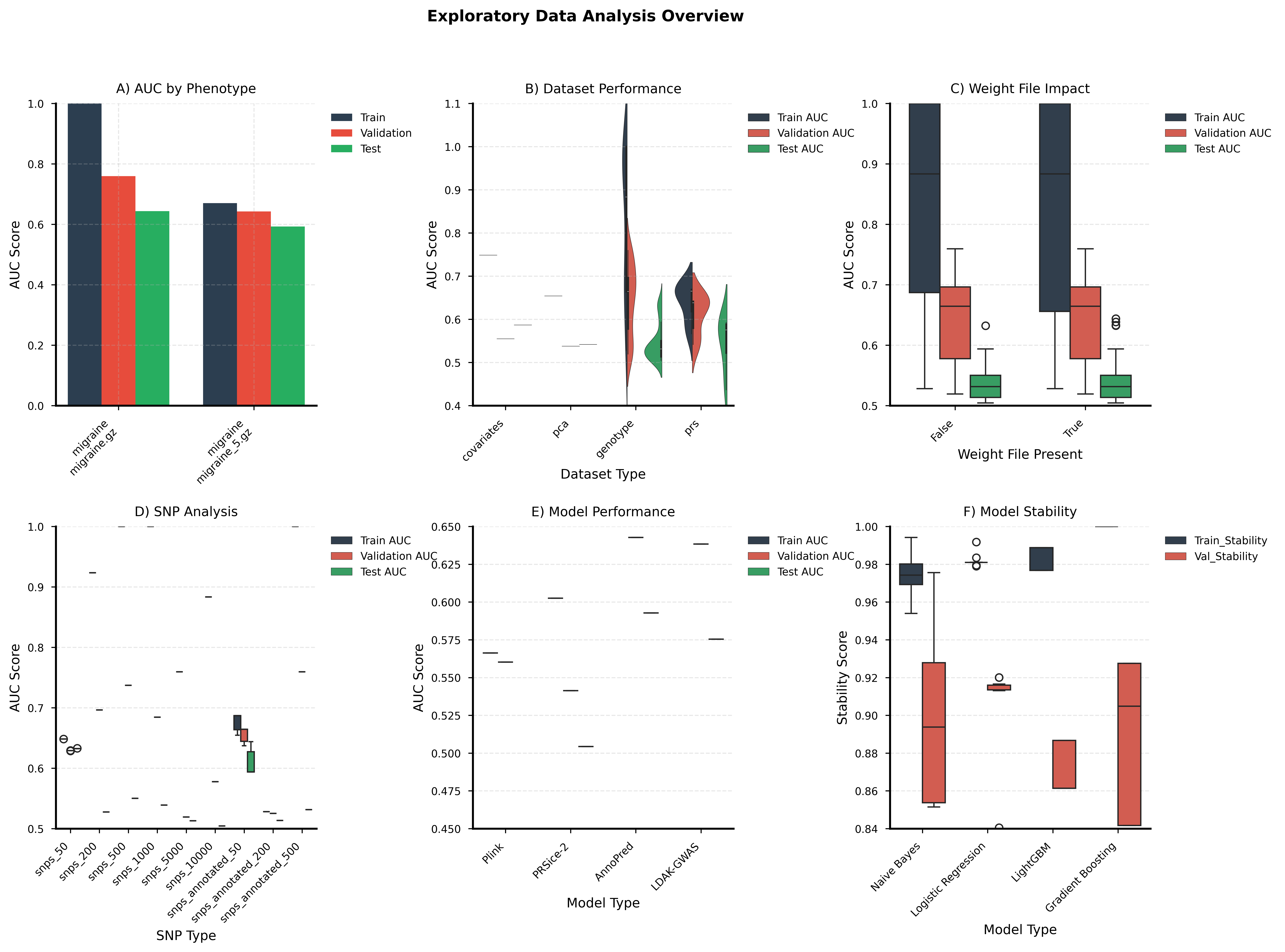}
\caption{\textbf{Impact of data-generation parameters on training, validation, and test performance for Configuration~1.} (A) Best performance for each GWAS. (B) Comparative performance across dataset types, shown using violin plots of AUC scores for each dataset category. (C) Impact of weight-file incorporation on model performance. (D) Performance by number of SNPs. (E) Comparison of PRS model performance. (F) Stability analysis of machine-learning models across all datasets. Analyses were conducted using the top model for each dataset, selected by composite score.}
\label{EDA1}
\end{figure}

For Configuration~1, covariates and PCA were effectively fixed representations, whereas genotype-derived and PRS-derived datasets displayed meaningful variability across GWAS source, SNP threshold, weighting status, and model choice. Among the migraine-derived inputs, compact genotype representations based on 50 SNPs or 50 annotated SNPs were among the strongest retained candidates, and weighted versions improved discrimination in several settings. GWAS1 (\texttt{migraine.gz}) consistently outperformed GWAS2 (\texttt{migraine\_2.gz}), achieving training, validation, and test AUCs of approximately 0.95, 0.75, and 0.63, respectively. PRS-derived datasets showed more heterogeneous behaviour across tools, with PRSice-2 contributing competitively within the retained search space. Logistic regression demonstrated the highest stability across training and validation splits among the machine-learning models evaluated (Figure~\ref{EDA1}).

\begin{figure}[!ht]
\centering
\includegraphics[width=\textwidth]{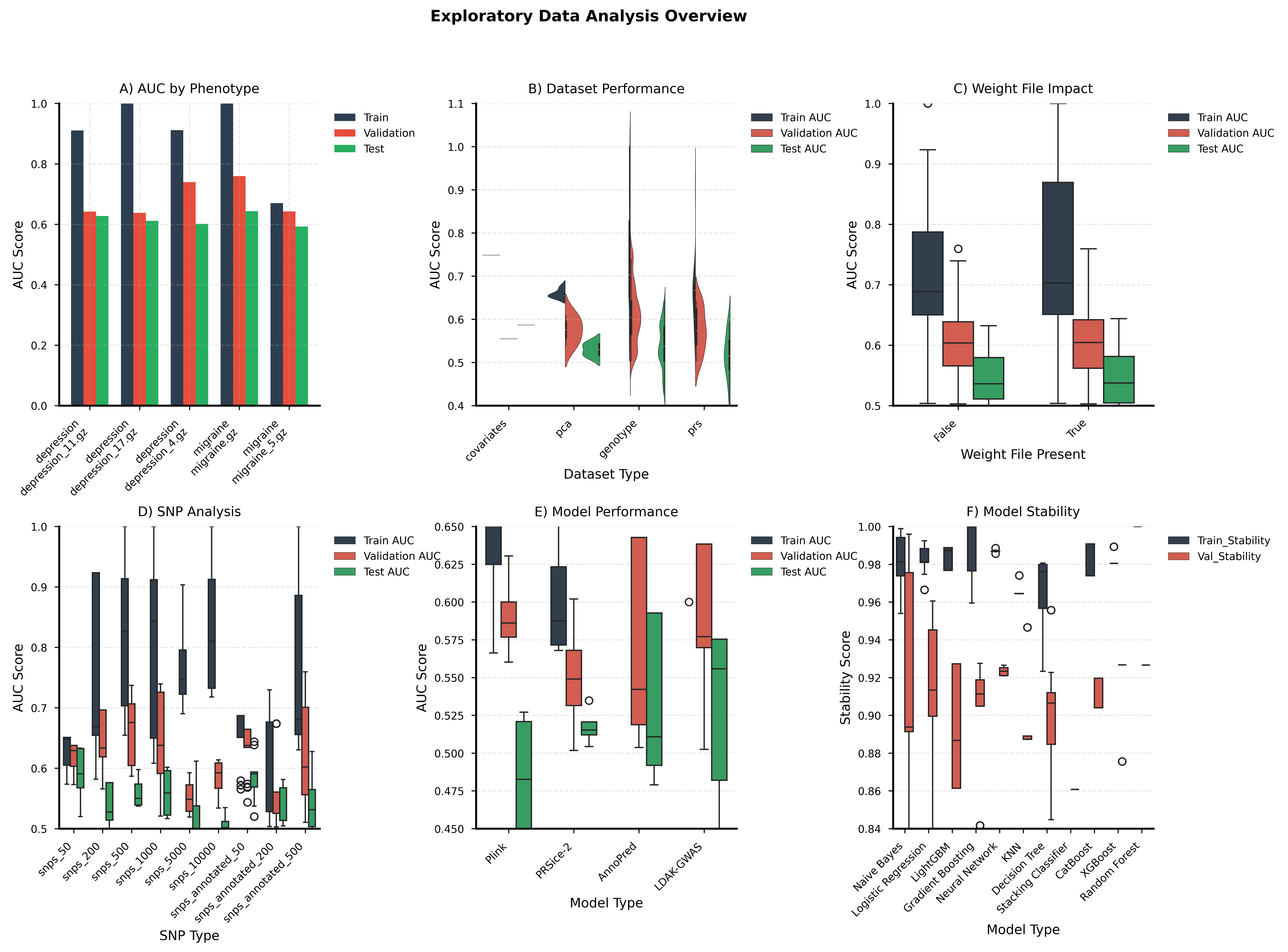}
\caption{\textbf{Impact of data-generation parameters on training, validation, and test performance for Configuration~2.} (A) Best performance across GWAS sources. (B) Comparative performance across dataset types. (C) Impact of weight-file incorporation on model performance. (D) Performance by number of SNPs. (E) Comparison of PRS model performance. (F) Stability analysis of machine-learning models across all datasets. Analyses were conducted using the top model for each dataset, selected by composite score.}
\label{EDA2}
\end{figure}

For Configuration~2, GWAS1 (\texttt{migraine.gz}) again achieved the strongest overall performance among the retained GWAS-specific representations (training: 0.95, validation: 0.75, test: 0.63), while the remaining migraine-derived and depression-derived inputs showed more comparable behaviour. Weighted genotype datasets generally improved discrimination, and compact representations based on 50 SNPs or 50 annotated SNPs showed comparable training and test performance. When using genotype data derived from depression GWAS, performance for annotated 50-SNP configurations decreased relative to the equivalent migraine-derived configuration, suggesting that annotation-based feature prioritization is more effective when the GWAS source is phenotype-matched. Selected depression-derived PRS representations remained competitive, suggesting that related-trait inputs may contribute useful signal when prioritized empirically rather than added indiscriminately. The neural network demonstrated the highest stability between training and validation sets among all machine-learning models evaluated in Configuration~2 (Figure~\ref{EDA2}).

\subsection*{PRS tool performance provides broader benchmarking context}

To contextualise the PRS results obtained within EFGPP, we draw on findings from a companion study~\cite{Muneeb2026prstools} in which 46 PRS tools were benchmarked across seven binary UK Biobank phenotypes under harmonised preprocessing and cross-validation conditions, using the same 733-participant cohort and covariate structure employed here. In that study, the best-performing tool for migraine was XPBLUP (full-model AUC $= 0.629$), followed by GEMMA-LMM (AUC $= 0.629$) and GEMMA-LM (AUC $= 0.628$), with PRSice-2 achieving AUC $= 0.549$ under the same full-model configuration. These values provide an externally validated single-tool ceiling for migraine PRS prediction in this cohort and confirm that the best individual PRS dataset within EFGPP (test AUC $= 0.521$ for PLINK-derived migraine PRS in the ablation analysis) is consistent with the range reported across tools in the broader benchmark.

Critically, the best multimodal combination in EFGPP Configuration~1 (test AUC $= 0.688$) exceeded the best single-tool full-model result from the 46-tool benchmark (AUC $= 0.629$) by $+0.059$ AUC units, demonstrating that structured multimodal integration provides a measurable improvement beyond what any single PRS tool can achieve in isolation, even after covariate adjustment. Depression-derived PRS contributed competitive predictive signal for migraine prediction across all four tools evaluated; among all PRS configurations, LDAK-GWAS-depression achieved the strongest result (AUC\,=\,0.564, 95\% CI [0.520, 0.608], $\Delta=-0.075$), outperforming every migraine-derived PRS, consistent with documented genetic overlap between the two traits~\cite{Ligthart2014}.

\subsection*{Structured multimodal integration improves migraine prediction}

Single-modality analysis showed that no genetic representation alone matched the covariates-only baseline, although genotype-derived matrices were consistently more competitive than PRS in isolation. Structured integration improved performance beyond these individual inputs. In the migraine-focused configuration, the best multimodal combination achieved a mean test AUC of 0.688, compared with 0.644 for the strongest individual representation and 0.639 for the covariates-only baseline. In the cross-trait configuration incorporating depression-derived inputs, the best multimodal model achieved a mean test AUC of 0.663. These results indicate that EFGPP can identify complementary data representations that improve prediction under limited-sample conditions, while also showing that related-trait data must be empirically prioritized rather than added indiscriminately.

We created dataset combinations and trained each using the stacked artificial neural network framework described in STAR Methods. For Configuration~1, 11 selected datasets yielded 1{,}012 candidate combinations of 2--5 datasets before filtering; after discarding combinations with mean validation AUC below 0.50, 427 unique combinations remained. For Configuration~2, 16 selected datasets yielded 6{,}868 candidate combinations before filtering, of which 366 passed the validation threshold. Results for all combinations are available on GitHub (\texttt{Configuration1/ResultsML1.csv}, \texttt{Configuration2/ResultsML2.csv}).

Systematic evaluation of multimodal combinations confirmed that structured integration of heterogeneous genetic representations improves migraine prediction beyond any single data modality (Tables~\ref{tab:config1_results}--\ref{tab:combinations}). In Configuration~1, which restricted inputs to migraine-derived sources, two three-dataset combinations tied for the highest mean test AUC. The most parsimonious was the combination of covariates, PCA, and a weighted non-annotated migraine genotype matrix (mean test AUC $0.688 \pm 0.109$, 95\% CI $[0.552, 0.824]$, $\Delta = +0.049$); a second combination of covariates and two weighted non-annotated migraine genotype matrices at different SNP thresholds achieved a comparable test AUC of $0.685 \pm 0.127$ (95\% CI $[0.527, 0.843]$, $\Delta = +0.046$). A five-dataset combination of covariates, PCA, and three genotype matrices reached $0.683 \pm 0.064$ with a narrower confidence interval, suggesting improved stability when additional complementary representations are included. Notably, no PRS dataset appeared in the top-performing Configuration~1 combinations; weighted non-annotated genotype matrices from migraine GWAS were the dominant genetic representation across all top models, suggesting that in this setting, directly modelled SNP-level genotype data captures more predictive signal than aggregated PRS scores. Across all top Configuration~1 combinations, covariates were consistently retained alongside genotype-derived features, reinforcing the conclusion that clinical and metabolomic covariates capture phenotype-proximal signal that complements purely genetic predictors.

\begin{table}[!ht]
\centering
\caption{\textbf{Top-performing dataset combinations for Configuration~1.} Configuration~1 used migraine GWAS inputs only. Size indicates the number of datasets in the combination. Train, validation, and test AUC are means across five cross-validation folds. The top-performing combinations all included weighted or unweighted non-annotated genotype matrices from migraine GWAS, with covariates consistently retained in the highest-ranking models.}
\label{tab:config1_results}
\small
\resizebox{\textwidth}{!}{%
\begin{tabular}{cp{0.62\linewidth}ccc}
\toprule
\textbf{Size} & \textbf{Datasets} & \textbf{Train AUC} & \textbf{Val AUC} & \textbf{Test AUC} \\
\midrule
3 & Covariates $+$ PCA $+$ Genotype\_W\_not\_annotated\_migraine & 0.90 & 0.66 & 0.69 \\
3 & Covariates $+$ Genotype\_W\_not\_annotated\_migraine $\times$2 & 0.95 & 0.74 & 0.69 \\
5 & Covariates $+$ PCA $+$ Genotype\_W\_not\_annotated\_migraine $\times$2 $+$ Genotype\_UW\_not\_annotated\_migraine & 0.78 & 0.67 & 0.68 \\
4 & Covariates $+$ PCA $+$ Genotype\_UW\_not\_annotated\_migraine $+$ Genotype\_W\_not\_annotated\_migraine & 0.91 & 0.69 & 0.67 \\
2 & Genotype\_W\_not\_annotated\_migraine $\times$2 & 0.73 & 0.70 & 0.66 \\
\bottomrule
\end{tabular}%
}
\end{table}

In Configuration~2, which incorporated depression-derived inputs alongside migraine representations, the best combination was a two-dataset model comprising unweighted annotated migraine genotype data and a depression-derived LDAK-GWAS PRS score, achieving a test AUC of $0.663 \pm 0.048$ (95\% CI $[0.603, 0.723]$, $\Delta = +0.024$). A second two-dataset combination using \texttt{Genotype\_UW\_annotated} and \texttt{PRS\_depression\_1\_AnnoPred} achieved a test AUC of 0.658 ($\pm$ 0.063). The five-dataset combination (\texttt{PRS\_depression\_2\_PLINK}, \texttt{PRS\_migraine\_LDAK-GWAS}, \texttt{PRS\_depression\_1\_AnnoPred}, \texttt{Genotype\_W\_not\_annotated}, and \texttt{Genotype\_UW\_annotated}) achieved a training AUC of 0.980 ($\pm$ 0.011) but a test AUC of only 0.623 ($\pm$ 0.058), suggesting overfitting when multiple datasets are combined without sufficient regularization. A notable pattern is that combinations including both depression-related PRS scores (\texttt{PRS\_depression\_2\_PLINK} and \texttt{PRS\_depression\_1\_AnnoPred}) tend to achieve higher training and validation scores but show slightly reduced test performance. The consistent presence of \texttt{Genotype\_UW\_annotated} across all top-performing Configuration~2 combinations highlights its central role, while the competitive performance of depression-derived inputs is consistent with documented genetic overlap between the two traits~\cite{Ligthart2014}.

\begin{table}[!ht]
\centering
\caption{\textbf{Top-performing dataset combinations for Configuration~2.} Configuration~2 used migraine and depression GWAS inputs. Size indicates the number of datasets in the combination. Train, validation, and test AUC are means across five cross-validation folds.}
\label{tab:config2_results}
\small
\resizebox{\textwidth}{!}{%
\begin{tabular}{cp{0.66\linewidth}ccc}
\toprule
\textbf{Size} & \textbf{Datasets} & \textbf{Train AUC} & \textbf{Val AUC} & \textbf{Test AUC} \\
\midrule
2 & Genotype\_UW\_annotated $+$ PRS\_migraine\_2\_LDAK-GWAS & 0.80 & 0.71 & 0.66 \\
2 & Genotype\_UW\_annotated $+$ PRS\_depression\_1\_AnnoPred & 0.87 & 0.74 & 0.66 \\
4 & PRS\_depression\_2\_PLINK $+$ PRS\_depression\_1\_PLINK $+$ Genotype\_UW\_not\_annotated $+$ Genotype\_UW\_annotated & 0.90 & 0.72 & 0.63 \\
5 & PRS\_depression\_2\_PLINK $+$ PRS\_migraine\_LDAK-GWAS $+$ PRS\_depression\_1\_AnnoPred $+$ Genotype\_W\_not\_annotated $+$ Genotype\_UW\_annotated & 0.98 & 0.80 & 0.62 \\
3 & PRS\_depression\_2\_PLINK $+$ Genotype\_UW\_annotated $+$ Genotype\_W\_not\_annotated & 0.85 & 0.78 & 0.62 \\
\bottomrule
\end{tabular}%
}
\end{table}

\begin{table}[!ht]
\centering
\caption{\textbf{Top-performing multimodal combinations for migraine prediction.} Results are reported as mean test AUC $\pm$ standard deviation across five cross-validation folds, with 95\% confidence intervals (CI) computed using the $t$-distribution with $n-1$ degrees of freedom. $\Delta$ denotes the improvement in test AUC relative to the covariates-only baseline (test AUC $= 0.639$). Configuration~1 uses migraine GWAS inputs only; Configuration~2 additionally incorporates depression-derived PRS and genotype representations. UW = unweighted; W = weighted; Ann = functionally annotated; NoAnn = not annotated.}
\label{tab:combinations}
\scriptsize
\resizebox{\textwidth}{!}{%
\begin{tabular}{cp{0.52\linewidth}cccccc}
\toprule
\textbf{Config} & \textbf{Datasets} & \textbf{Size} & \textbf{Train AUC} & \textbf{Val AUC} & \textbf{Test AUC} & \textbf{95\% CI} & \textbf{$\Delta$} \\
\midrule
Config~1 & Covariates $+$ PCA $+$ Genotype (W, NoAnn, migraine) & 3 & 0.902 & 0.655 & $0.688 \pm 0.109$ & [0.552, 0.824] & $+0.049$ \\
Config~1 & Covariates $+$ Genotype (W, NoAnn, migraine) $\times$2 & 3 & 0.953 & 0.740 & $0.685 \pm 0.127$ & [0.527, 0.843] & $+0.046$ \\
Config~1 & Covariates $+$ PCA $+$ Genotype (W, NoAnn, migraine) $\times$2 $+$ Genotype (UW, NoAnn, migraine) & 5 & 0.780 & 0.670 & $0.683 \pm 0.064$ & [0.604, 0.762] & $+0.044$ \\
Config~1 & Covariates $+$ PCA $+$ Genotype (UW, NoAnn, migraine) $+$ Genotype (W, NoAnn, migraine) & 4 & 0.911 & 0.688 & $0.668 \pm 0.117$ & [0.523, 0.814] & $+0.029$ \\
Config~1 & Genotype (W, NoAnn, migraine) $\times$2 & 2 & 0.727 & 0.695 & $0.658 \pm 0.125$ & [0.503, 0.814] & $+0.019$ \\
\midrule
Config~2 & Genotype (UW, Ann, migraine) $+$ PRS\_migraine\_2\_LDAK-GWAS & 2 & 0.798 & 0.714 & $0.663 \pm 0.048$ & [0.603, 0.723] & $+0.024$ \\
Config~2 & Genotype (UW, Ann, migraine) $+$ PRS\_depression\_1\_AnnoPred & 2 & 0.870 & 0.735 & $0.658 \pm 0.063$ & [0.580, 0.736] & $+0.019$ \\
Config~2 & PRS\_depression\_2\_PLINK $+$ PRS\_depression\_1\_PLINK $+$ Genotype (UW, NoAnn, depression) $+$ Genotype (UW, Ann, migraine) & 4 & 0.903 & 0.722 & $0.634 \pm 0.140$ & [0.460, 0.808] & $-0.005$ \\
Config~2 & PRS\_depression\_2\_PLINK $+$ PRS\_migraine\_LDAK-GWAS $+$ PRS\_depression\_1\_AnnoPred $+$ Genotype (W, NoAnn) $+$ Genotype (UW, Ann) & 5 & 0.980 & 0.803 & $0.623 \pm 0.058$ & [0.551, 0.695] & $-0.016$ \\
Config~2 & PRS\_depression\_2\_PLINK $+$ Genotype (UW, Ann, migraine) $+$ Genotype (W, NoAnn, depression) & 3 & 0.849 & 0.785 & $0.620 \pm 0.107$ & [0.487, 0.753] & $-0.019$ \\
\bottomrule
\end{tabular}%
}
\end{table}

\subsection*{Cross-trait inputs provide signal but do not automatically improve integration}

The observation that depression-derived PRS and genotype inputs contributed to competitive migraine prediction models in Configuration~2 is consistent with documented genetic overlap between the two traits~\cite{Ligthart2014}, and suggests that cross-disorder representations can add predictive value when empirically prioritised within a structured framework rather than added indiscriminately. Depression-derived PRS performed comparably to or better than migraine-derived PRS across all four tools in the single-modality analysis, with LDAK-GWAS-depression achieving the strongest result among all PRS configurations.

However, the performance improvements from multimodal integration were modest in absolute terms, with the best combination improving upon the covariates-only baseline by $+0.049$ AUC units in Configuration~1 and $+0.024$ in Configuration~2. The best cross-trait multimodal configuration reached a mean test AUC of 0.663, which was competitive but lower than the best migraine-focused multimodal configuration (AUC\,=\,0.688). This pattern is consistent with the incremental and complementary nature of the genetic signal being integrated and underscores the framework's purpose as a tool for systematic data prioritisation rather than as a method for producing large absolute performance gains. These results indicate that related-trait inputs can contain useful predictive information, but their inclusion should be empirically prioritized rather than assumed to improve performance.

\subsection*{EFGPP provides a reproducible framework for future phenotype-scale benchmarking}

The EFGPP workflow is designed to support reproducible representation prioritization across heterogeneous genotype--phenotype inputs. Code and workflow documentation are provided through the project repository, and the analysis explicitly separates dataset generation, model fitting, ranking, pruning, multimodal selection, and held-out evaluation within each fold. Although the present study focuses on migraine prediction, the same framework can be extended to additional phenotypes, larger cohorts, external validation datasets, and broader benchmarking studies. This positions EFGPP as a reusable decision framework for phenotype-scale evaluation of genotype-derived, PRS-derived, covariate, and cross-trait representations.

\section*{DISCUSSION}

This study presents \textsc{EFGPP} as a practical framework for prioritizing and integrating heterogeneous inputs for genotype--phenotype prediction. Rather than introducing a new PRS algorithm, the framework addresses a common but under-served modelling problem: when multiple GWAS, PRS methods, genotype representations, annotations, and covariates are available, which inputs are most informative to carry forward, and which combinations generalize best? In this proof-of-concept application to migraine prediction, \textsc{EFGPP} identified a best individual representation with a test AUC of 0.644 and improved performance to 0.688 through structured multimodal integration in a migraine-focused configuration, with a cross-trait configuration reaching 0.663. Framed in this way, the main contribution of the study is methodological and practical: it offers a reproducible strategy for reducing a large space of plausible data representations into a smaller set of better-supported modelling choices.

Several findings from the single-dataset and integrated analyses are particularly informative. First, no single genetic representation dominated across all settings. The covariates-only model established a strong non-genetic baseline (AUC 0.639), and no single genetic modality exceeded it. Among the single-modality genetic representations, genotype-based features were more competitive than PRS, with the strongest depression-derived genotype representation reaching AUC 0.620 and the best migraine-derived genotype representations reaching AUC 0.605. By contrast, the strongest migraine-derived PRS reached AUC 0.521, while the strongest overall PRS representation was depression-derived \textsc{LDAK-GWAS} at AUC 0.564. These findings suggest that, in this cohort, directly modelled SNP-level genotype information retained more predictive signal than aggregated PRS scores when each modality was evaluated in isolation.

Second, the top multimodal models were generally parsimonious, and adding more datasets did not guarantee improved generalization. In Configuration~1, the best-performing model combined only three inputs -- covariates, PCA, and a weighted non-annotated migraine genotype matrix -- and achieved the strongest overall performance (AUC 0.688). A slightly more complex five-dataset model showed somewhat narrower variability but did not surpass this mean test AUC. This is an important result in its own right: in modestly sized cohorts, disciplined feature prioritization may matter more than raw feature volume. It also explains why no PRS representation appeared in the strongest Configuration~1 combinations, despite PRS remaining biologically relevant and competitive in some single-modality settings.

The cross-trait analyses provide an additional biologically and methodologically relevant signal. Migraine and depression are well known to co-occur, and accumulating evidence supports partial overlap in their biology and genetic architecture~\cite{Ligthart2014}. Within this context, the observation that depression-derived PRS contributed to competitive migraine prediction models is plausible and strengthens the rationale for testing related-trait information within a structured framework. In particular, the best Configuration~2 model combined migraine-derived genotype features with depression-derived \textsc{LDAK-GWAS} PRS and achieved AUC 0.663. Importantly, these results should not be interpreted as showing that cross-disorder data are universally beneficial; rather, they suggest that related-trait inputs can be useful when they are empirically prioritized and selectively integrated. That distinction matters, because naive feature expansion can increase apparent training performance without improving out-of-sample prediction.

The broader benchmarking context is also informative. In a companion harmonized study of 46 PRS tools evaluated in the same 733-participant UK Biobank cohort, the best single-tool full-model migraine AUC was 0.629. Against that benchmark, the best multimodal Configuration~1 result in \textsc{EFGPP} (AUC 0.688) represents a gain of $+0.059$ AUC units, indicating that structured multimodal integration can improve upon the ceiling achieved by individual PRS tools alone. At the same time, the absolute gains observed here remain modest: relative to the covariates-only baseline, the best multimodal model improved test AUC by $+0.049$ in Configuration~1 and $+0.024$ in Configuration~2. This pattern is consistent with the incremental and complementary nature of the signal being integrated and underscores that the value of \textsc{EFGPP} lies primarily in systematic data prioritization rather than in producing large absolute performance gains.

The current study should therefore be interpreted as a proof-of-concept rather than a clinically deployable prediction model. The sample size is modest, the performance gains are incremental, and the strongest integrated models include non-genetic covariates that may capture phenotype-proximal or comorbidity-related information in addition to inherited risk. Accordingly, the present study is best viewed as demonstrating a workflow for representation selection and multimodal benchmarking, not as establishing a migraine risk predictor ready for translational use. Stronger claims would require larger cohorts, broader benchmarking against contemporary PRS methods, more formal nested model selection, calibration analyses, and external validation in independent datasets.

Several limitations define the next stage of development. The framework explores a large combinatorial space, and although ranking and category-based reduction make this tractable, the search process remains computationally intensive. Performance may depend on the available discovery GWAS, the ancestry composition of both discovery and target data, the selected SNP thresholds, and the specific feature categories retained after prioritization. More broadly, PRS and related genetic predictors remain sensitive to discovery sample composition and often show reduced portability across settings, emphasizing the need for cautious interpretation and external validation. Future work should therefore focus on extending the framework across phenotypes, ancestries, and larger cohorts; incorporating stronger contemporary PRS baselines; formalizing nested selection; and improving reproducibility through clearer reporting of each decision point in the prioritization pipeline.

Taken together, our results support a simple conclusion: for complex trait prediction, the key challenge is often not data scarcity, but data prioritization. \textsc{EFGPP} provides a structured way to evaluate multiple GWAS-derived and genotype-derived representations, identify those with the strongest evidence of utility and stability, and test whether they contribute complementary predictive information when combined. In migraine, this strategy improved performance beyond any single representation and showed that selectively integrated cross-trait information can add value. More broadly, the framework offers a practical foundation for future studies seeking to move from opportunistic feature assembly toward systematic, evidence-based construction of genotype--phenotype prediction models.

\section*{Limitations of the study}

This study has several limitations. First, the analysis was performed in a modestly sized UK Biobank cohort of 733 individuals, including 53 migraine cases and 680 controls, which limits the precision and generalizability of the performance estimates. Second, external validation in an independent cohort was not performed, and the results should therefore be interpreted as proof of concept rather than evidence of clinical deployability. Third, calibration analysis was not performed, and the framework was not evaluated as a clinical risk prediction model. Fourth, although the fold-wise design reserved held-out test subsets for final evaluation, the large representation search remains computationally intensive and should be further evaluated in larger cohorts. Fifth, the analysis focused on migraine and depression-derived cross-trait inputs; future work should test additional phenotypes, ancestries, PRS tools, and external datasets.

\newpage

 \section*{RESOURCE AVAILABILITY}

\subsection*{Lead contact}

Requests for further information and resources should be directed to and will be fulfilled by the lead contact, David B. Ascher (d.ascher@uq.edu.au).

\subsection*{Materials availability}

This study did not generate new physical materials or reagents.

\subsection*{Data and code availability}

\begin{itemize}
    \item UK Biobank individual-level genotype, phenotype, covariate, and metabolomic data were accessed under UK Biobank application ID 50000. These data are subject to UK Biobank access restrictions and cannot be redistributed by the authors. Researchers may apply for access through the UK Biobank Access Management System: \url{https://www.ukbiobank.ac.uk/}.
    
    \item GWAS summary statistics used in this study were obtained from the GWAS Catalog. The accession identifiers are GCST90038646, GCST90043745, GCST90038650, GCST90101808, and GCST005839.
    
    \item Functional annotation files were obtained from AnnoPred: \url{https://github.com/yiminghu/AnnoPred}.
    
    \item Linkage disequilibrium reference data were obtained from LDAK: \url{https://dougspeed.com/reference-panel/}.
    
    \item The EFGPP framework source code and workflow documentation are available at: \url{https://github.com/MuhammadMuneeb007/EFGPP}.
    
    \item Additional processed outputs and configuration files required to reproduce the reported analyses are available through the EFGPP repository or from the lead contact upon reasonable request, subject to UK Biobank data access restrictions.
\end{itemize}

\section*{ACKNOWLEDGMENTS}

D.B.A. is supported by an NHMRC Investigator Grant (GNT2041888).

\section*{AUTHOR CONTRIBUTIONS}

Conceptualization, M.M. and D.B.A.; methodology, M.M. and D.B.A.; investigation, M.M.; software, M.M.; formal analysis, M.M.; writing--original draft, M.M.; writing--review and editing, M.M. and D.B.A.; funding acquisition, D.B.A.; resources, M.M. and D.B.A.; supervision, D.B.A.

\section*{DECLARATION OF INTERESTS}
The authors declare no competing interests.

\section*{DECLARATION OF GENERATIVE AI AND AI-ASSISTED TECHNOLOGIES}

Claude and GitHub Copilot were used to support code review, language editing, and workflow figure preparation. The authors reviewed and edited all outputs as needed and take full responsibility for the content of this publication.


\section*{STAR METHODS}

\subsection*{Key resources table}

\begin{table}[!ht]
\centering
\caption{\textbf{Key resources used in this study.}}
\label{tab:key_resources}
\small
\begin{tabular}{p{0.27\linewidth}p{0.33\linewidth}p{0.32\linewidth}}
\toprule
\textbf{Resource} & \textbf{Source} & \textbf{Identifier / link} \\
\midrule
UK Biobank genotype, phenotype, covariate, and metabolomic data & UK Biobank & Application ID 50000; \url{https://www.ukbiobank.ac.uk/} \\
GWAS Catalog & European Bioinformatics Institute & \url{https://www.ebi.ac.uk/gwas/} \\
Migraine GWAS summary statistics & GWAS Catalog & GCST90038646; GCST90043745 \\
Depression GWAS summary statistics & GWAS Catalog & GCST90038650; GCST90101808; GCST005839 \\
PLINK & PLINK & \url{https://www.cog-genomics.org/plink/} \\
PRSice-2 & PRSice-2 & \url{https://www.prsice.info/} \\
AnnoPred & AnnoPred & \url{https://github.com/yiminghu/AnnoPred} \\
LDAK-GWAS & LDAK & \url{https://dougspeed.com/} \\
LDAK linkage disequilibrium reference data & LDAK & \url{https://dougspeed.com/reference-panel/} \\
EFGPP source code and workflow documentation & GitHub & \url{https://github.com/MuhammadMuneeb007/EFGPP} \\
Python & Python Software Foundation & \url{https://www.python.org/} \\
scikit-learn & Python package & \url{https://scikit-learn.org/} \\
TensorFlow/Keras & Python package & \url{https://www.tensorflow.org/} \\
XGBoost & Python package & \url{https://xgboost.readthedocs.io/} \\
LightGBM & Python package & \url{https://lightgbm.readthedocs.io/} \\
CatBoost & Python package & \url{https://catboost.ai/} \\
\bottomrule
\end{tabular}
\end{table}

\subsection*{Experimental model and study participant details}

This study used individual-level genotype, phenotype, covariate, and metabolomic data from UK Biobank under application ID 50000. The analysis focused on migraine case--control prediction. The final study cohort comprised 733 individuals, including 53 migraine cases and 680 controls. UK Biobank received ethical approval from the North West Multi-centre Research Ethics Committee, and all participants provided written informed consent.

This study did not involve recruitment of new participants, collection of new human samples, generation of new biological materials, or experimental intervention. All analyses were performed using previously collected UK Biobank data and publicly available GWAS-derived resources.

\subsection*{Method details}

\subsubsection*{EFGPP framework overview}

\textsc{EFGPP} is a reproducible framework for generating, prioritizing, pruning, and integrating heterogeneous genotype--phenotype data representations. The framework was designed to address a practical model-development question: given multiple GWAS, PRS tools, genotype representations, functional annotations, principal components, and covariates, which data representations and combinations should be carried forward for phenotype prediction under limited-sample conditions?

The workflow comprised six main stages: heterogeneous data-source assembly, candidate dataset generation, individual dataset benchmarking, redundancy pruning and category-level ranking, multimodal combination testing, and final held-out evaluation. Candidate representations included covariates, principal components, genotype-derived SNP matrices, GWAS-weighted SNP matrices, functionally annotated SNP matrices, weighted and annotated SNP matrices, and PRS-derived representations.

Six base data sources were used to construct the candidate representation space: GWAS summary statistics, target-cohort genotype data, phenotype labels, covariates, linkage disequilibrium reference data, and functional annotations. GWAS summary statistics were obtained from the GWAS Catalog. Genotype, phenotype, covariate, and metabolomic data were obtained from UK Biobank. Linkage disequilibrium reference data were obtained from LDAK, and functional annotation files were obtained from AnnoPred. PRS representations were generated using PLINK, PRSice-2, AnnoPred, and LDAK-GWAS. These inputs were processed into covariate, PCA, genotype-derived, annotated, weighted, weighted-and-annotated, and PRS-based dataset families before downstream benchmarking, pruning, ranking, and multimodal integration.

For each GWAS source, genotype-derived datasets were constructed by selecting SNPs at predefined thresholds and representing them as unweighted, GWAS-weighted, functionally annotated, or weighted-and-annotated matrices. PCA representations were used to capture population-structure-related variation, while covariate representations included clinical and metabolomic variables. PRS-derived representations were generated separately for each GWAS--tool combination. These candidate datasets formed the input space for downstream benchmarking, pruning, ranking, and multimodal integration.

All data-dependent processing steps were performed within the fold-wise evaluation design. Training and validation subsets were used for dataset construction, model fitting, hyperparameter comparison, ranking, pruning, and multimodal combination selection. Held-out test subsets were reserved exclusively for final fold-level performance evaluation and were not used to guide dataset selection, model choice, or combination search.

\subsubsection*{Phenotype definition and covariates}

Migraine case--control status was defined using UK Biobank phenotype information. The final analysis cohort contained 53 migraine cases and 680 controls. Covariate representations included 135 NMR metabolomic biomarkers and comorbid condition indicators. These covariates were used both as an individual non-genetic baseline and as candidate inputs for multimodal integration.

\subsubsection*{GWAS summary statistics}

Five GWAS summary statistic files were used to construct phenotype-matched and cross-trait representations. Two GWAS files were migraine-derived and three were depression-derived. Migraine-derived GWAS were used to construct phenotype-matched genetic representations, while depression-derived GWAS were included to evaluate whether cross-trait genetic signal could contribute to migraine prediction. The GWAS summary statistics were obtained from the GWAS Catalog and corresponded to accessions GCST90038646, GCST90043745, GCST90038650, GCST90101808, and GCST005839~\cite{D_nerta__2021,Jiang_2021,2018,Harder_2022}. GWAS files were quality controlled by retaining SNPs with minor allele frequency (MAF) $> 0.01$ and imputation information score (INFO) $> 0.8$, while removing ambiguous and missing variants~\cite{Turner2011,Truong2022}.  

\subsubsection*{Genotype-derived representations}

Genotype-derived datasets were constructed from UK Biobank genotype data using GWAS-informed SNP selection. Genotype quality control was applied using thresholds of MAF $> 0.01$, Hardy--Weinberg equilibrium $P > 1 \times 10^{-6}$, genotype missingness $\leq 0.1$, individual missingness $\leq 0.1$, and relatedness cutoff $\leq 0.125$~\cite{Anderson2010,Pavan2020}. SNPs were selected at predefined thresholds and converted into genotype matrices using $P$-value thresholding~\cite{Zeng2023}.

For weighted genotype matrices, SNP dosages were weighted using GWAS-derived effect estimates. For annotated matrices, selected SNPs were intersected with functional annotation information obtained from AnnoPred. These procedures generated compact and expanded genotype-derived representations across GWAS sources, SNP thresholds, weighting status, and annotation status.

\subsubsection*{Principal component analysis}

Principal component analysis was used to generate PCA-based representations from genotype data. PCA representations were included to account for population-structure-related variation and to test whether broad genetic structure alone contributed predictive information for migraine status.

\subsubsection*{Polygenic risk score generation}

PRS representations were generated using four PRS tools: PLINK, PRSice-2, AnnoPred, and LDAK-GWAS. These tools were selected to represent distinct PRS method families and were interpreted in the context of our companion harmonized benchmark of 46 PRS tools across UK Biobank phenotypes~\cite{Muneeb2026prstools}. PRS were generated separately for each GWAS source and tool combination. Migraine-derived PRS represented phenotype-matched genetic risk, while depression-derived PRS represented cross-trait genetic signal. These PRS-derived datasets were evaluated both as individual representations and as candidate inputs for multimodal integration.

\subsubsection*{Configuration design}

Two main configurations were evaluated. Configuration~1 focused on migraine-derived inputs and represented the migraine-focused search space. Configuration~2 expanded the search space by incorporating depression-derived GWAS and PRS inputs to evaluate cross-trait contribution.

In Configuration~1, 174 candidate datasets were generated and reduced to 62 non-redundant representations after similarity-based pruning. In Configuration~2, 975 candidate datasets were generated and reduced to 489 non-redundant representations. Following fold-wise benchmarking, validation-based filtering, and category-level ranking, 11 datasets from Configuration~1 and 16 datasets from Configuration~2 were selected for multimodal combination testing.

\subsubsection*{Fold-wise evaluation design}

The full dataset was partitioned into five stratified folds while preserving the migraine case--control ratio. For each fold, training, validation, and held-out test subsets were defined. Training and validation subsets were used for dataset construction, model fitting, hyperparameter comparison, dataset ranking, pruning, and multimodal combination selection. The held-out test subset for each fold was reserved exclusively for final fold-level evaluation.

To reduce information leakage during the large-scale search procedure, all data-dependent steps were performed within each fold using only the training and validation subsets. These steps included GWAS filtering, genotype quality control, SNP selection, functional annotation, PCA derivation, PRS generation, model fitting, hyperparameter comparison, dataset ranking, pruning, and multimodal combination selection. The held-out test subset was used only once for final fold-level evaluation after all within-fold selection decisions had been completed, and no test-set information was used to guide dataset prioritization, model choice, or combination search.

This design should be interpreted as a stratified five-fold held-out validation framework with within-fold train/validation/test splitting, rather than strict nested cross-validation.

\subsubsection*{Individual dataset benchmarking}

Each retained individual dataset was evaluated using 12 classification machine-learning algorithms. These included standard classifiers, ensemble models, gradient-boosting methods, and neural-network-based models. Models included logistic regression, decision tree, random forest, support vector machine, Naive Bayes, k-nearest neighbours, gradient boosting, XGBoost, LightGBM, CatBoost, multilayer perceptron, and artificial neural network classifiers.

Models were optimized via hyperparameter tuning where applicable, configured to handle class imbalance, and evaluated using training, validation, and held-out test performance across folds. For each dataset, the top-performing model was selected based on validation performance and the composite score. Datasets with mean validation AUC greater than 0.6 were retained for downstream prioritization and multimodal combination testing. Using this criterion, 49 datasets passed the validation threshold in Configuration~1 and 355 datasets passed the validation threshold in Configuration~2.

\subsubsection*{Redundancy pruning}

To reduce redundancy in the representation space, candidate datasets were compared using similarity-based pruning based on the Kolmogorov--Smirnov statistic~\cite{Massey1951}. Highly similar datasets were removed before downstream ranking and multimodal testing. This step was used to make the large representation space tractable while preserving diversity across major representation families, including covariates, PCA, genotype-derived features, and PRS-derived features.

\subsubsection*{Category-level ranking and composite score}

Retained datasets were grouped into categories to preserve representation diversity while reducing the multimodal search space. Categories included covariates, PCA, genotype-derived representations stratified by weighting and annotation status, and PRS representations stratified by GWAS file and PRS tool.

The retained datasets within each category were ranked using a composite score that incorporated validation AUC, train-validation generalization gap, training stability, and validation stability:

\begin{align}
\text{Composite score} =\;&
0.25 \times \text{Validation AUC}_{\text{norm}} \nonumber \\
&+ 0.25 \times \text{Train--validation gap}_{\text{norm}} \nonumber \\
&+ 0.25 \times \text{Train stability}_{\text{norm}} \nonumber \\
&+ 0.25 \times \text{Validation stability}_{\text{norm}}.
\end{align}

Top-ranked representatives from each category were carried forward for multimodal combination testing. This category-based approach was used to avoid selecting many near-duplicate datasets from the same representation family and to preserve modality diversity during integration.

\subsubsection*{Single-modality ablation analysis}

Single-modality ablation analysis was performed using individual dataset results generated within the EFGPP framework. For each of the five cross-validation folds, the best-performing machine-learning model was selected per dataset based on validation AUC.

Individual datasets were grouped into mutually exclusive categories according to data type, GWAS source phenotype, PRS tool, and genotype annotation and weighting status. For each category, the best-performing dataset was identified by mean test AUC across folds. Performance was summarized using mean training AUC, validation AUC, test AUC, standard deviation, 95\% confidence interval, and the difference in test AUC relative to the covariates-only baseline.

\subsubsection*{Multimodal integration}

Multimodal integration was performed after category-level prioritization. Selected datasets were combined systematically to test whether integrated representations improved prediction beyond individual modalities. For Configuration~1, 11 selected datasets were used to generate candidate combinations of 2--5 datasets. For Configuration~2, 16 selected datasets were used to generate candidate combinations of 2--5 datasets.

Each dataset combination was evaluated using a stacked artificial neural network framework. In this framework, each input dataset was processed as a separate representation branch, and learned representations were combined before final classification. Candidate combinations with low validation performance were filtered before final comparison. The final multimodal models were evaluated on held-out test subsets across folds.

For Configuration~1, 1{,}012 candidate combinations were generated before filtering, and 427 unique combinations remained after discarding combinations with mean validation AUC below 0.50. For Configuration~2, 6{,}868 candidate combinations were generated before filtering, and 366 passed the validation threshold. Results for all multimodal combinations are available in the EFGPP GitHub repository.

\subsubsection*{Cross-trait analysis}

Cross-trait analysis was performed by incorporating depression-derived GWAS and PRS representations into Configuration~2. These inputs were evaluated to determine whether related-trait genetic signal improved migraine prediction. Depression-derived PRS and genotype-derived representations were assessed both as individual representations and as components of multimodal combinations.

The cross-trait analysis was interpreted as a test of representation prioritization rather than a claim that related-trait inputs should always improve prediction. Depression-derived PRS showed evidence of transferable signal, but the best cross-trait multimodal configuration did not exceed the best migraine-focused multimodal configuration, indicating that related-trait data must be empirically prioritized rather than added indiscriminately.

\subsubsection*{PRS benchmarking context}

PRS results from EFGPP were interpreted alongside a companion benchmark of 46 PRS tools across seven binary UK Biobank phenotypes using harmonized preprocessing and cross-validation conditions. This comparison was used to contextualize the magnitude of PRS-only performance in the same cohort and to determine whether structured multimodal integration exceeded the performance range observed for individual PRS tools.

\subsection*{Quantification and statistical analysis}

Predictive performance was quantified using area under the receiver operating characteristic curve (AUC). For each dataset or model combination, training AUC, validation AUC, and held-out test AUC were calculated across five folds. Mean performance and standard deviation were reported across folds.

For single-modality ablation analysis, 95\% confidence intervals were calculated using the $t$-distribution across the five fold-level test AUC values:

\begin{equation}
CI = \bar{x} \pm t_{0.975, n-1} \times \frac{s}{\sqrt{n}},
\end{equation}

where $\bar{x}$ is the mean test AUC, $s$ is the standard deviation of test AUC across folds, and $n=5$ is the number of folds.

The performance difference relative to the covariates-only baseline was calculated as:

\begin{equation}
\Delta_{\text{covariates}} = \text{AUC}_{\text{model}} - \text{AUC}_{\text{covariates}}.
\end{equation}

For multimodal integration, candidate combinations were ranked primarily by mean held-out test AUC, while validation performance was used during within-fold model selection and filtering. Final claims were based on held-out test performance aggregated across folds.

No external validation cohort was used. Calibration analysis was not performed. The study should therefore be interpreted as a proof-of-concept evaluation of a representation-prioritization and multimodal benchmarking framework, rather than as a clinically deployable migraine risk prediction model.

\bibliography{references}

\end{document}